\documentclass[aps,pre,showpacs,showkeys,preprint,eqsecnum,floatfix,superscriptaddress,nofootinbib]{revtex4-1}

\usepackage{amssymb}
\usepackage{amsmath}
\usepackage{amsbsy}
\usepackage{bm}
\usepackage{color}
\usepackage{graphicx}
\usepackage[normalem]{ulem} 
\usepackage{calrsfs}

\newcommand{\bq}{\begin{eqnarray}}
\newcommand{\eq}{\end{eqnarray}}
\newcommand{\bqn}{\begin{eqnarray*}}
\newcommand{\eqn}{\end{eqnarray*}}

\newcommand{\rr}{\mathbf{r}}

\newcommand{\nablab}{\pmb{\nabla}}

\newcommand{\calp}{{\cal P}}
\newcommand{\calh}{{\cal H}}
\newcommand{\calt}{{\cal T}}
\newcommand{\calv}{{\cal V}}
\newcommand{\cala}{{\cal A}}
\newcommand{\calo}{{\cal O}}

\begin{document}
%%%%%%%%%%%%%%%%%%%%%%%%%%%%%%%%%%%%%%%%%%%%%%%%%%%%%%%%%%%%%%%%%%%%%%%%%%%%%%
%%%%%%%%%%%%%%%%%%%%%%%%%%%%%%%%%%%%%%%%%%%%%%%%%%%%%%%%%%%%%%%%%%%%%%%%%%%%%%
%%%%%%%%%%%%%%%%%%%%%%%%%%%%%%%%%%%%%%%%%%%%%%%%%%%%%%%%%%%%%%%%%%%%%%%%%%%%%%
\title{Two component boson-fermion plasma at finite temperature} 

\author{Riccardo Fantoni}
\email{rfantoni@ts.infn.it}
\affiliation{Universit\`a di Trieste, Dipartimento di Fisica, strada
  Costiera 11, 34151 Grignano (Trieste), Italy}
\date{\today}

\begin{abstract}
We discuss thermodynamic stability of neutral real (quantum) matter
from the point of view of a computer experiment at finite, non-zero,
temperature. We perform (restricted) path integral Monte Carlo
simulations of the two component plasma where the two species are all
bosons, all fermions, and one boson and one fermion. We calculate
the structure of the plasma and discuss about the formation of binded
couples of oppositely charged particles. The purely bosonic case is
thermodynamically unstable. In this case we find an undetermined size
dependent contact value of the unlike partial radial distribution
function. For the purely fermionic case we find a demixing transition
with binding also of like species. 
\end{abstract}



\maketitle
%%%%%%%%%%%%%%%%%%%%%%%%%%%%%%%%%%%%%%%%%%%%%%%%%%%%%%%%%%%%%%%%%%%%%%%%%%%%%%
\section{Introduction}
%%%%%%%%%%%%%%%%%%%%%%%%%%%%%%%%%%%%%%%%%%%%%%%%%%%%%%%%%%%%%%%%%%%%%%%%%%%%%%
\label{sec:introduction}

For matter to be stable it must be globally neutral.
It is well known that in order for a system of an equal number $N$ of
oppositely charged point particles to be stable against collapse
quantum mechanics is required, and furthermore at least one of the
species of particles must be a fermion. Without the exclusion
principle the ground state energy per particle of the system diverges
as $N^{7/5}$ and the thermodynamic limit is not well defined
\cite{Lieb1976}. As a matter of fact in the classical limit one is
forced to introduce a short range regularization (like an hard core or
others) \cite{Fantoni16b} of the pair-potential between the particles
in order to prevent the collapse of the negative charges on the
positive ones. \cite{Fantoni13e,Fantoni13f} All this is at the heart
of the fundamental question of whether the matter we live in is stable
or not.

In this work we want to explore the structure of a two component
mixture of particles of two opposite charge species. We will consider
particles of charge $\pm e$ with $e$ the charge of an
electron. Furthermore we will assume that the two species both have
the mass of an electron $m$. We will consider explicitly the cases
where both species have spin $1/2$ (purely fermionic), when they both
have spin 1 (purely bosonic) and when one species has spin $1/2$ and
one species has spin 1 (fermions-bosons mixture). In all
cases we assume that each species has polarization equal to 1.
Doing so we will be able to determine the thermodynamic instability of
the purely bosonic case as opposed to the other two cases. We will
work at high temperature and intermediate density, when the quantum
effects are not very important. The path integral Monte Carlo computer
experiment is only exact in the purely bosonic case apart from the
usual finite size and imaginary time discretization errors. For the
other two cases it is necessary to resort to an approximation due to
the {\sl fermions sign problem}. \cite{Ceperley1991,Ceperley1996} We
will choose the restricted path integral approximation with a
restriction based on the nodes of the ideal density matrix,
which is known to perform reasonably well for the one component
(Jellium) case from the pioneering work of Brown {\sl et al.}
\cite{Brown2013,Brown2014}. Other methods has been 
implemented recently in order to reach high densities: Bonitz {\sl et
  al.} \cite{Dornheim2016,Groth2017} combine configuration path
integral Monte Carlo and permutation blocking path integral Monte
Carlo. Malone {\sl et al.} \cite{Malone2016} agrees well with the one
of Bonitz at high densities and the direct path integral Monte Carlo
one of Filinov {\sl et al.} \cite{Filinov2015} that agrees well with
Brown at low density and moderate temperature. Our method is
alternative to all previously employed ones.

In our simulations we use the {\sl worm algorithm}
\cite{Boninsegni2006a,Boninsegni2006b} which is able to sample the
necessary permutations of the indistinguishable particles without the
need of explicitly sampling the permutations space
treating the paths as ``worms'' with a tail ({\sl Masha}) and a head
({\sl Ira}) in the $\beta$-periodic imaginary time, which can be
attached one with the other in different ways or swap some of their
portions. We explicitly and efficiently applied the restriction to the
worms and this allowed us to treat the fermionic or mixed case
explicitly albeit only approximately. The approximation is expected to
become better at low density and high temperature, i.e. when
correlation effects are weak.

Possible physical realizations of interest to our work for the case of
both species of spin $1/2$ are a
non-relativistic electron-positron plasmas created in the laboratory
\cite{Iwamoto1993} or an electron-hole plasma 
which is important in the realm of low-temperature semiconductor
physics. Conduction electrons and holes in semiconductors interact
with Coulomb force and can have very similar effective masses.
\cite{Wolfe1995,Shumway1999}
 
The work is organized as follows: In section \ref{sec:model} we
describe the physical model we want to study, in section
\ref{sec:simulation} we describe the computer experiment method and
techniques, in section \ref{sec:results} we describe our numerical
results, and section \ref{sec:conclusions} is for final remarks.

%%%%%%%%%%%%%%%%%%%%%%%%%%%%%%%%%%%%%%%%%%%%%%%%%%%%%%%%%%%%%%%%%%%%%%%%%%%%%%
\section{The model}
%%%%%%%%%%%%%%%%%%%%%%%%%%%%%%%%%%%%%%%%%%%%%%%%%%%%%%%%%%%%%%%%%%%%%%%%%%%%%%
\label{sec:model}

Setting lengths in units of the Bohr radius $a_0=\hbar^2/me^2$ and
energies in Rydberg's units, $\text{Ry}=\hbar^2/2ma_0^2$, where $m$ is
the electron mass, the Hamiltonian of the two component non-relativistic
electron-positron mixture is 
\bq 
\calh&=&\calt+\calv=-\lambda\sum_{i=1}^{N_+}\nablab_{\rr^+_i}^2-
\lambda\sum_{i=1}^{N_-}\nablab_{\rr^-_i}^2+V(R)~,\\ 
V&=&2\left(\sum^{N_+}_{i<j}\frac{1}{|\rr^+_i-\rr^+_j|}+ 
\sum^{N_-}_{i<j}\frac{1}{|\rr^-_i-\rr^-_j|}-
\sum^{N_+}_{i=1}\sum^{N_-}_{j=1}\frac{1}{|\rr^+_i-\rr^-_j|}\right)~,
\eq
where $\lambda=\hbar^2/2ma_0^2=\text{Ry}$,
$R=(\rr^+_1,\ldots,\rr^+_{N_+},\rr^-_1,\ldots,\rr^-_{N_-})$ with 
$\rr^+_i$ the coordinates of the $i$th positron and $\rr^-_i$ the ones
of the $i$th electron. We will choose $N_+=N_-=N$ since the system
must be neutrally charged in order to be thermodynamically stable.
We will not introduce any short range regularization of the Coulomb
potential. And we will treat the Coulomb long range potential using
the Ewald sums technique \cite{Natoli1995} in order to treat it in the
periodic box of side $L$ of the simulation.

We will treat explicitly the electron-positron case where the two
particles are both fermions, the case where both species are bosons,
and the case where one species only is a fermion. Of course there is
no charged boson in nature with the mass and the charge of the
electron, so this will remain a speculative analysis, to explore
the thermodynamic stability and statistical properties of the mixture.

We will carry on a grand canonical simulation at fixed chemical
potentials of the two species $\mu^+,\mu^-$, volume $\Omega=L^3,$ and
absolute temperature $T=1/k_B\beta$, with $k_B$ the Boltzmann
constant. 
   
%%%%%%%%%%%%%%%%%%%%%%%%%%%%%%%%%%%%%%%%%%%%%%%%%%%%%%%%%%%%%%%%%%%%%%%%%%%%%%
\section{Simulation method}
%%%%%%%%%%%%%%%%%%%%%%%%%%%%%%%%%%%%%%%%%%%%%%%%%%%%%%%%%%%%%%%%%%%%%%%%%%%%%%
\label{sec:simulation}

We carry on a (restricted) path integral Monte Carlo computer
experiment \cite{Ceperley1995} using the {\sl worm algorithm}
\cite{Boninsegni2006a,Boninsegni2006b} to simulate the behavior of the
quantum mixture at finite temperature.  
 
The {\sl density matrix} of a system of many distinguishable bodies at
temperature $k_BT=\beta^{-1}$ can be written as an integral over all
paths $\{R_t\}$
\bq
\rho(R_\beta,R_0;\beta)=\oint_{R_0\to R_\beta}dR_t\,\exp(-S[R_t]).
\eq
the path $R_t$ begins at $R_0$ and ends at $R_\beta$. For
non-relativistic particles interacting with a potential $V(R)$ the {\sl
  action} of the path, $S[R_t]$, is given by the Feynman-Kac formula
\bq
S[R_t]=\int_0^\beta dt\,\left[\frac{1}{4\lambda}\left|\frac{dR_t}{dt}\right|^2
+V(R_t)\right].
\eq 
Thermodynamic properties, such as the radial distribution function
(RDF), are related to the diagonal part of the density matrix, so that
the path returns to its starting place after a time $\beta$. 

To perform Monte Carlo calculations of the integrand, one makes
imaginary thermal time discrete with a {\sl time step} $\tau$, so that
one has a finite (and hopefully small) number of time slices and thus
a classical system of $N$ particles in $M=\beta/\tau$ time
slices; an equivalent $NM$ particle classical system of ``polymers''.
\cite{Ceperley1995} 

Thermodynamic properties are averages over the thermal $2N-$body
density matrix which is defined as a thermal occupation of the exact
eigenstates $\phi_i(R)$
\bq
\rho(R,R';\beta)=\sum_i\phi_i^*(R)e^{-\beta E_i}\phi_i(R').
\eq
The partition function is the trace of the density matrix
\bq
Z(\beta)=e^{-\beta F}=\int dR\,\rho(R,R;\beta)=\sum_ie^{-\beta E_i},
\eq
with $F$ Helmholtz's free energy. Other thermodynamic averages are
obtained as 
\bq
\langle\calo\rangle=Z(\beta)^{-1}\int dR dR'\,\langle R|\calo|R'\rangle
\rho(R',R;\beta).
\eq

Path integrals are constructed using the product property of density
matrices
\bq
\rho(R_2,R_0;\beta_1+\beta_2)=\int dR_1\,
\rho(R_2,R_1;\beta_2)\rho(R_1,R_0;\beta_1),
\eq
which holds for any sort of density matrix. If the product property is
used $M$ times we can relate the density matrix at a temperature
$\beta^{-1}$ to the density matrix at a temperature $M\beta^{-1}$. The
sequence of intermediate points $\{R_1,R_2,\ldots,R_{M-1}\}$ is the
path, and the {\sl time step} is $\tau=\beta/M$. As the time step gets
sufficiently small the Trotter theorem tells us that we can assume
that the kinetic ${\cal T}$ and potential ${\cal V}$ operator commute
so that: $e^{-\tau\calh}=e^{-\tau{\cal T}}e^{-\tau{\cal V}}$ (strictly
speaking this is only possible when ${\cal V}$ is bounded from below
\cite{Simon1979} but this is always satisfied by our simulation since
we use a radial discretization of the pair Coulomb potential) and the
{\sl primitive approximation} for the boltzmannon density matrix is
found \cite{Ceperley1995} 
\bq
\rho(R_0,R_M;\beta)&=&\int dR_1\ldots dR_{M-1}\,
\exp\left[-\sum_{m=1}^M S^m\right],\\
K^m&=&\frac{3N}{2}\ln(4\pi\lambda\tau)+\frac{(R_{m-1}-R_m)^2}{4\lambda\tau},\\
S^m-K^m&\approx& U^m_\text{primitive}=\frac{\tau}{2}[V(R_{m-1})+V(R_m)].
\eq 
The Feynman-Kac formula for the
boltzmannon density matrix results from taking the limit $M\to\infty$.
The price we have to pay for having an explicit expression for the
density matrix is additional integrations; all together
$3N(M-1)$. Without techniques for multidimensional integration,
nothing would have been gained by expanding the density matrix into a
path. Fortunately, simulation methods can accurately treat such
integrands. It is feasible to make $M$ rather large, say in the
hundreds or thousands, and thereby systematically reduce the time-step
error. The leading error of the primitive approximation goes like
$\sim \lambda\tau^2$. \cite{Ceperley1995}

In addition to sampling the path one also needs to sample all the
various necessary permutations of the indistinguishable particles
(bosons or fermions) and this is accomplished on the fly through the
use of the worm algorithm. \cite{Boninsegni2006a,Boninsegni2006b}

When we are dealing with bosons or fermions
$\rho_{B,F}(R_\beta,R_0;\beta)=\cala_\calp\rho(R_\beta,\calp 
R_0;\beta)$ is the density matrix corresponding to some set of quantum
numbers which is obtained by using the projection operator
$\cala_\calp=\frac{1}{N!}\sum_\calp(\pm)^\calp$, where $\calp$ is a
permutation of particles labels and the permutation sign is a plus for
bosons (B) and a minus for fermions (F), on the distinguishable particle
density matrix. Then for bosons we can carry on the Monte Carlo
calculation without further approximations, but for fermions the
following {\sl Restricted Path Integral} approximation is also
necessary in order to overcome the ubiquitous sign problem
\cite{Ceperley1991,Ceperley1996}   
\bq \label{rpii}
\rho_F(R_\beta,R_0;\beta)=\int dR'\,\rho_F(R',R_0;0)
\oint_{R'\to R_\beta\in\gamma_T(R_0)}dR_t\,e^{-S[R_t]},
\eq
where the subscript means that we restrict the path integration to
paths starting at $R'$, ending at $R_\beta$ and avoiding the nodes
(the zeroes) of a known {\sl trial density matrix}, $\rho_T$, assumed
to have nodes, $\partial\gamma_T$, close to the true ones. The weight
of the walk is $\rho_F(R',R_0;0)=(N!)^{-1}\sum_\calp(-)^\calp\delta(R'-\calp
R_0)$. It is clear that the contribution
of all the paths for a single element of the density matrix will be of
the same sign, thus avoiding the sign problem. On the
diagonal the density matrix is positive and on the path restriction
$\rho_F(R,R_0;\beta)>0$ then only even permutations are allowed
since $\rho_F(R,\calp R;\beta)=(-)^\calp\rho_F(R,R;\beta)$. It is then
possible to use a bosonic calculation to get the approximate fermionic
case. 

The restriction is implemented choosing as the trial density matrix
the ideal density matrix: we just reject the move ({\sl remove}, {\sl
  close}, {\sl wiggle}, and {\sl displace} in the Z sector, and {\sl
  advance} and {\sl swap} in the G sector)
\cite{Boninsegni2006a,Boninsegni2006b} whenever the proposed path is
such that the ideal fermionic or 
fermionic-bosonic density matrix calculated between the reference
point and any of the time slices subject to newly generated particles
positions has a negative value. 

The ideal fermionic or fermionic-bosonic density matrix is given by
\bq
\rho_0(R,R';t)\propto\cala \left(\begin{array}{cc}
e^{-\frac{(\rr^+_i-\rr^{+\prime}_j)^2}{4\lambda t}} & 
e^{-\frac{(\rr^+_i-\rr^{-\prime}_k)^2}{4\lambda t}}\\
e^{-\frac{(\rr^-_l-\rr^{+\prime}_j)^2}{4\lambda t}} & 
e^{-\frac{(\rr^-_l-\rr^{-\prime}_k)^2}{4\lambda t}}
\end{array}\right),
\eq
where $\lambda=\hbar^2/2m$ and $\cala$ is the (anti)symmetrization
operator for the positive and negative species (purely fermionic
mixture) or for the positive species only (fermionic-bosonic
mixture). We expect this approximation to be best at 
high temperatures and low densities when the correlation (the
particles coupling and their quantum nature) effects are
weak. Clearly in a simulation of the ideal gas ($V=0$) this
restriction returns the exact result for fermions, otherwise it is
just an approximation.

The restriction or the fixed nodes path integral may have an influence
on the thermodynamic stability of the fluid under study expecially at
low temperatures when quantum effects becomes more relevant. On the
other hand If this were the case it would have an influence on the
stability of the fluid under all thermodynamic states which we can
clearly exclude since as soon as we include at least one fermionic
species in the binary mixture the system becomes thermodynamically
stable even at moderately low temperatures when the restriction is not
very effective.  

%%%%%%%%%%%%%%%%%%%%%%%%%%%%%%%%%%%%%%%%%%%%%%%%%%%%%%%%%%%%%%%%%%%%%%%%%%%%%%
\section{Results}
%%%%%%%%%%%%%%%%%%%%%%%%%%%%%%%%%%%%%%%%%%%%%%%%%%%%%%%%%%%%%%%%%%%%%%%%%%%%%%
\label{sec:results}

In our simulations we chose $k_BT=10\text{Ry}$ and $L=5a_0$. Going
to lower temperatures the contact value for the unlike partial RDF
tends to increase since the binding between a positive and a negative
charge increases. This is because the coupling constant of the mixture
is $\Gamma=\beta e^2/a_0$. For the purely bosonic case the contact
value never reaches an equilibrium during the simulation evolution
unlike for the purely fermionic case or the fermions-bosons mixture
where a positive charge binds with a negative charge in a stable way
at low densities. \cite{Pierleoni1994}

It is also useful to introduce a degeneracy temperature $\Theta=T/T_F$,
where $T_F=T_D2\pi^2/\alpha_3^{2/3}$ is the Fermi temperature, here
$\alpha_3=4\pi/3$, and  
\bq
T_D=\frac{2n^{2/3}}{k_B}\text{Ry},
\eq
with $n=Na_0^3/V$ the density, is the degeneracy temperature. For
temperatures higher than $T_D$, as in our simulations, quantum effects
are less relevant. For this reason we chose $M=10$ in all cases giving
a $\tau=0.01\text{Ry}^{-1}$. So the primitive approximation is a good
one. 

Another relevant parameter is the Wigner-Seitz radius $r_s=(3/4\pi
n)^{1/3}$ which in the degenerate regime $\Theta\ll 1$ regulates
whether the system of particles is dominated by the potential energy
or by the kinetic energy. At high $r_s$ the potential energy dominates
and the system tends to crystallize. \cite{Shumway1999}

\begin{figure}[htbp]
\begin{center}
\includegraphics[width=10cm]{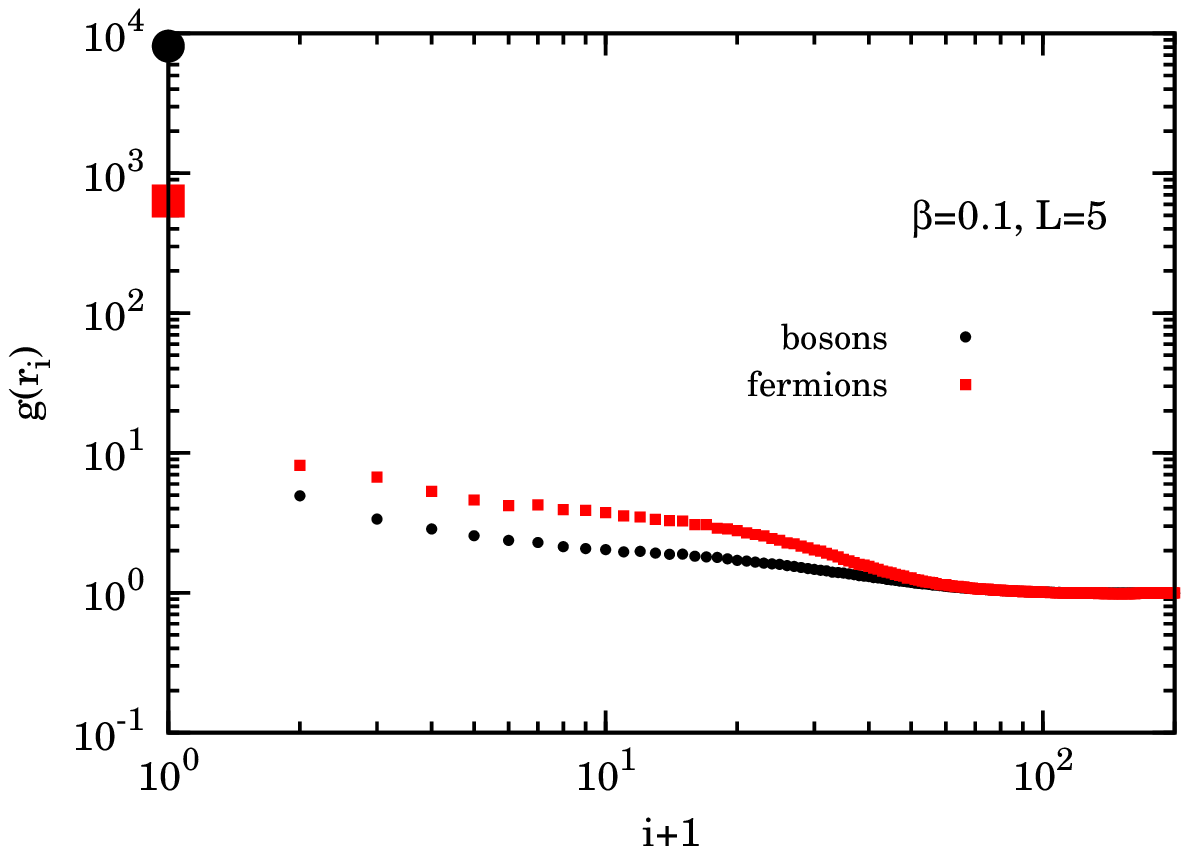}\\
\includegraphics[width=10cm]{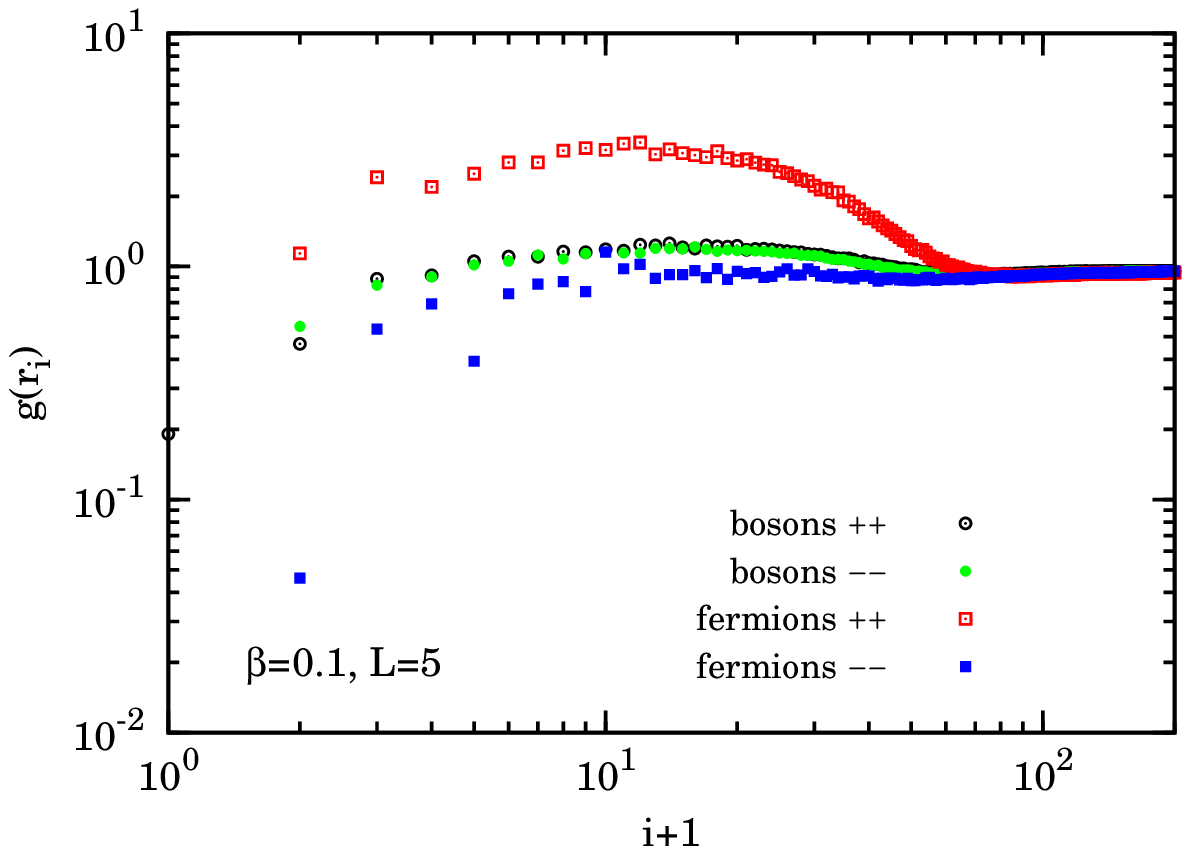}\\
\end{center}  
\caption{We show the partial RDF on a
  log-log scale. For the mixture of bosons and the fluid with one
  bosonic species and one fermionic species, we show $g_{+-}(r_i)$ in
  the upper panel and $g_{++}(r_i),g_{--}(r_i)$ in the bottom
  panel. In all cases we have $L/2=r_\text{cut}a_0=2.5a_0$ and the RDF
  are calculated on 200 radial points $r_i=idr$
  with $dr=r_\text{cut}/200$. The simulation was carried on at
  $\beta=0.1\text{Ry}^{-1}$ with $M=10$ time slices and an average of
  approximately 36 particles for the fermions case and 39 for the
  bosons case. The simulation was 15000 blocks of 500 steps taking
  averages every 100 moves. But $g_{+-}(0)$ for the purely bosonic
  case continued to grow afterwards.}   
\label{fig:gr1}
\end{figure}

\begin{figure}[htbp]
\begin{center}
\includegraphics[width=10cm]{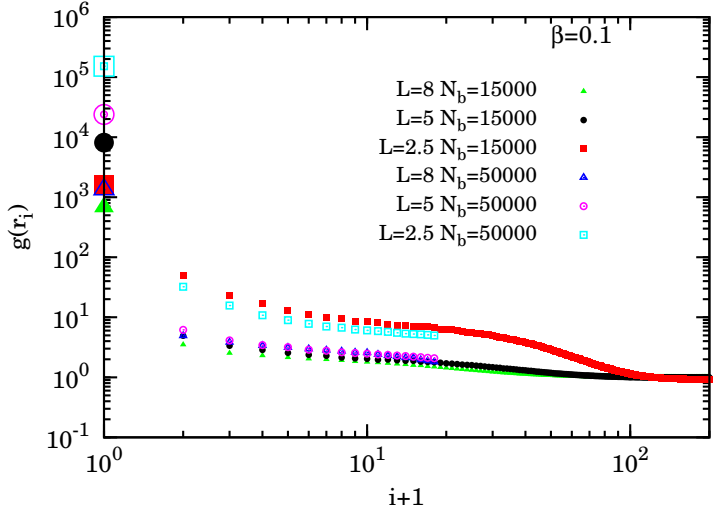}
\end{center}  
\caption{We show the unlike partial RDF on a log-log scale for the
  purely bosonic case at three different values of $L/2=r_\text{cut}a_0$
  and approximately same density and at two times during the
  simulation, after $N_b=15000$ blocks (of 50000 worm moves) and after
  $N_b=50000$ blocks. The RDF are calculated on 200 radial
  points $r_i=idr$ with $dr=r_\text{cut}/200$. The simulation was
  carried on at $\beta=0.1\text{Ry}^{-1}$ with $M=10$ time slices. The
  simulation was 15000 blocks of 500 steps taking averages every 100
  moves. But $g_{+-}(0)$ continued to grow afterwards.}  
\label{fig:gr2}
\end{figure}

From Fig. \ref{fig:gr1} we see how the binary mixture is stable when
the particles are fermions and unstable when they are bosons. This is
manifested by a contact value of the unlike partial RDF, for the purely
bosonic case, which is one order of magnitude higher than the one for
the purely fermionic case. It varies wildly during the simulation
evolution, with variations of one or more orders of magnitudes upon
inspections of the simulation at different time intervals of 10000
blocks of 50000 worm moves each. 
The like partial RDF for the purely fermionic case shows a
spontaneous symmetry braking where the positive-positive RDF differs
from the negative-negative one and presents a broad shoulder near the
origin which suggests the formation of like positive pairs.   
The contact value in the bosonic case has huge variations upon changes
of the size of the system as shown by Fig. \ref{fig:gr2}. This also
means that there is not a well defined thermodynamic limit of the RDF
which in turn is a manifestation of the system instability.
\cite{Lieb1976} This does not occur when at least one of the two
species is a fermion. In this case a slight shoulder near the origin
in the unlike partial RDF indicates the stable pairing between a
positive and a negative charge. The shoulder grows at lower
temperature and lower density.

\begin{figure}[htbp]
\begin{center}
\includegraphics[width=10cm]{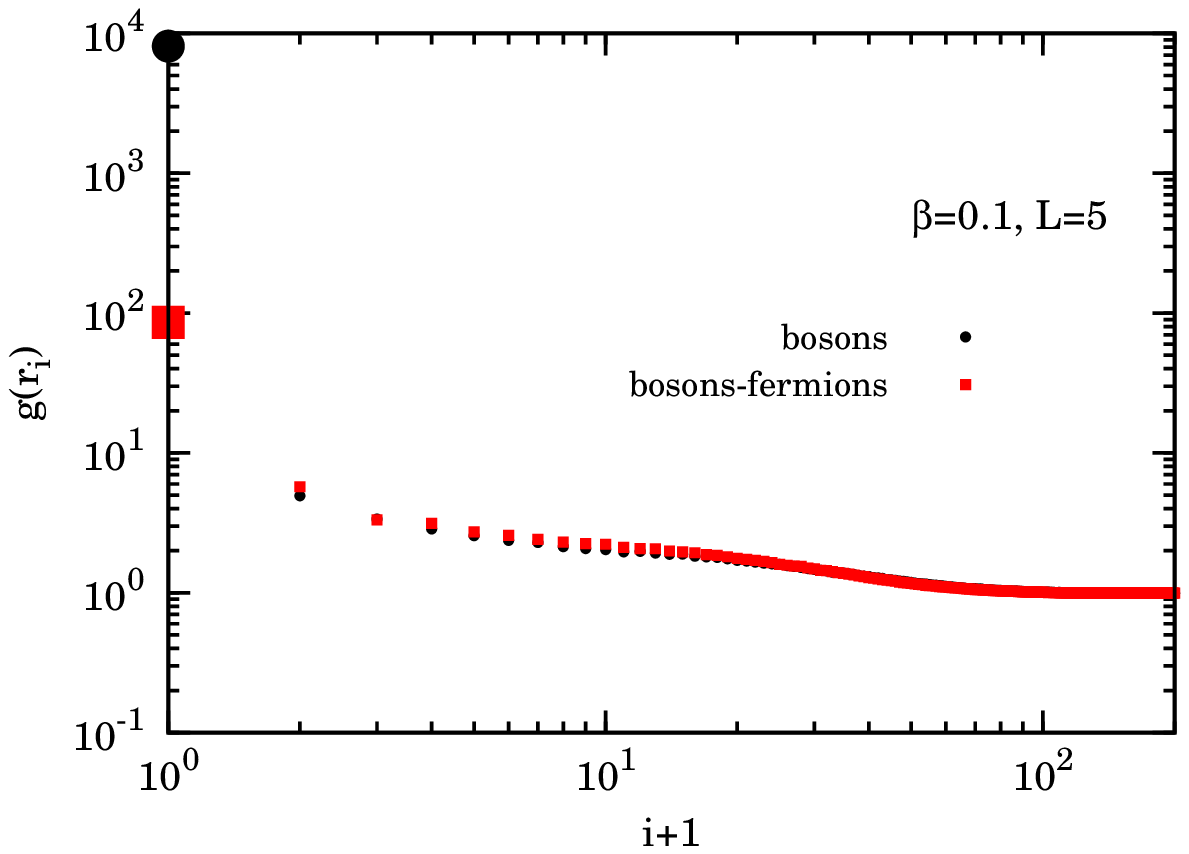}\\
\includegraphics[width=10cm]{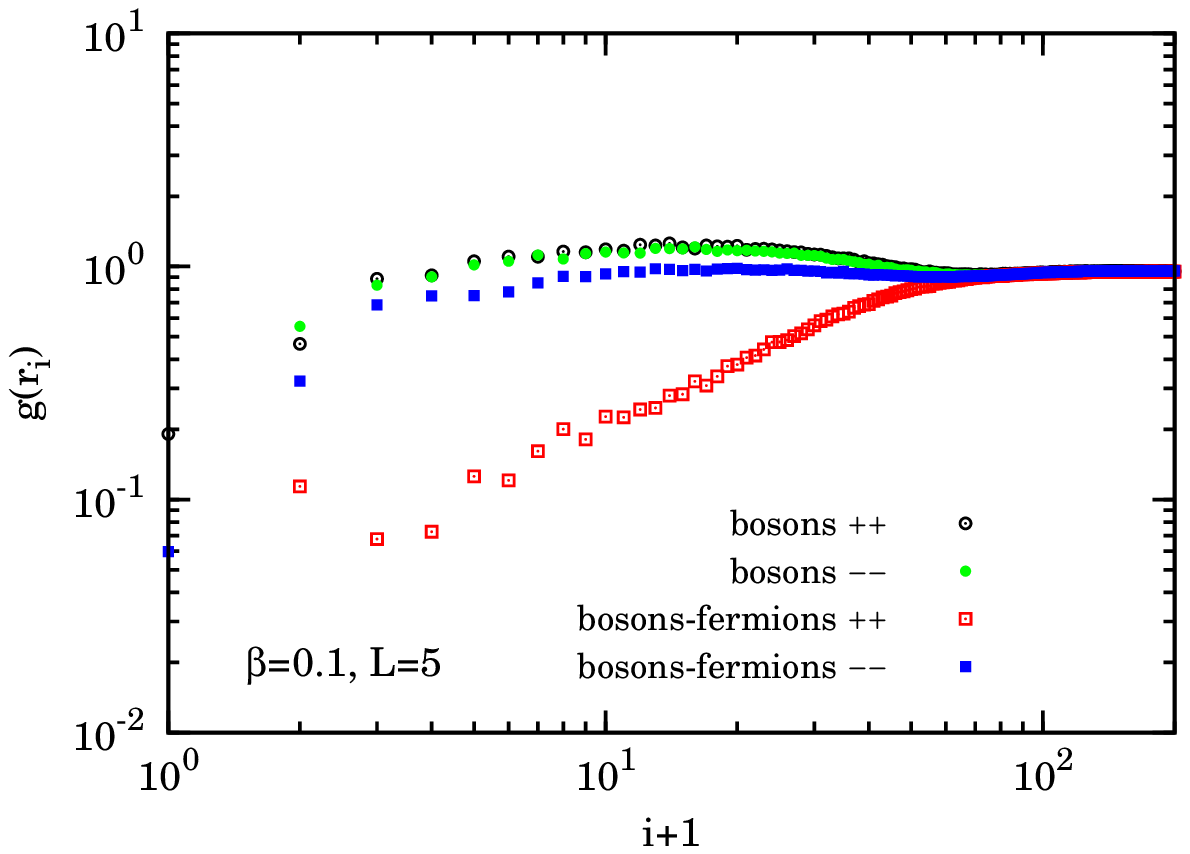}\\
\end{center}  
\caption{We show the partial RDF on a
  log-log scale. For the mixture of bosons and the fluid with one
  bosonic species and one fermionic species, we show $g_{+-}(r_i)$ in
  the upper panel and $g_{++}(r_i), g_{--}(r_i)$ in the bottom
  panel. In all cases we have $L/2=r_\text{cut}a_0=2.5a_0$ and the RDF
  are calculated on 200 radial points $r_i=idr$ with
  $dr=r_\text{cut}/200$. The simulation was carried on at
  $\beta=0.1\text{Ry}^{-1}$ with $M=10$ time slices and an average of
  approximately 38 particles for the mixed fermionic-bosonic case and
  39 for the purely bosonic case. The simulation was 15000 blocks of
  500 steps taking averages every 100 moves. But $g_{+-}(0)$ for the
  purely bosonic case continued to grow afterwards.}
\label{fig:gr3}
\end{figure}

In order to have stability it is sufficient to have at least one of
the two particle species to be a fermion as is shown in
Fig. \ref{fig:gr3}. In this case the like partial RDF for the bosonic
species is comparable with the one of the purely bosonic case and the
one for the fermionic component is below. No like pair formation is
visible from the structure analysis. The unlike partial RDF is
superposed to the one of the purely fermionic case but presents an on
top value two orders of magnitudes smaller.   

The difference between the purely fermionic mixture and the
fermions-bosons had to be expected also from the point of view of the
fact that our spin polarized fermions, unlike the bosons, do not have
a state with zero total angular momentum.

%

%%%%%%%%%%%%%%%%%%%%%%%%%%%%%%%%%%%%%%%%%%%%%%%%%%%%%%%%%%%%%%%%%%%%%%%%%%%%%%
\section{Conclusions}
%%%%%%%%%%%%%%%%%%%%%%%%%%%%%%%%%%%%%%%%%%%%%%%%%%%%%%%%%%%%%%%%%%%%%%%%%%%%%%
\label{sec:conclusions}

In conclusion we carried on some computer experiments for the binary
mixture of oppositely charged pointwise particle species when both
species are bosons, both fermions, and one bosons and one fermions. We
chose the charge and the mass equal to the ones of the electron and
only considered fully polarized species. We used the worm algorithm to
perform (restricted) path integral Monte Carlo simulations, at finite
temperatures. 

We simulated the mixture with a weak degree of degeneracy $\Theta\sim
1.4$ and a weak coupling $\Gamma=0.2$. The Wigner-Seitz radius for
each species was $r_s\sim 1$.

During the simulations we measured the radial distribution function of
the three mixtures and found that the purely bosonic one is
thermodynamically unstable toward the collapse of oppositely charged
particles ones upon the others. Whereas in the other two mixtures the
Pauli exclusion principle restores the stability producing stable
bindings: like pairs form for the purely fermionic case as a result
of a spontaneous symmetry breaking in a demixing transition and unlike
pairs form in both cases. 
The instability manifests itself through a pronounced peak in 
the contact value of the unlike partial RDF which is strongly size
dependent in the experiment and keeps growing as the simulation
evolves without ever reaching convergence towards a stable value. This
observation tells us that the fermionic character of the simplest
constituent of matter is essential in nature to be able to have a
stable matter. On the other hand if one uses non-quantum statistical
mechanics one must regularize the Coulomb potential at short range, for
example through the addition of an hard core to the otherwise pointwise
particles \cite{Fantoni13e,Fantoni13f}. Even if in the relativistic
regime it is plausible to 
talk about an electron radius, attempts to model the electron as a
non-point particle are considered ill-conceived and counter-pedagogic.
\cite{Curtis2003} 
 
In order to have a stable matter it is necessary that it is globally
neutral and that it is made up of at least one fermionic
species. Physical realizations of our model are non-relativistic
electron-positron plasma produced in the laboratory \cite{Iwamoto1993}
and electron-hole plasma in semiconductors \cite{Wolfe1995}. Of course
in the numerical experiment we do not have the physical limitations
that occur in a laboratory. This allowed us to inquire also the
mixture with one bosonic or even both bosonic components. Another
interesting issue where our study could become relevant is atom and
molecule formation. In its simplest setting this involves the 
study of an electron-proton mixture. Since the mass of an electron is
three orders of magnitude smaller that the one of the proton the degeneracy
temperature of the electron species is three orders of magnitude
smaller than the one of the nuclei, at a given density. Therefore it
is very unlikely that an electron, with a world-line with many
particle exchanges will bind to a nucleus, which has a world-line
with many less particle exchanges. In order for this to occur we have
to go down to temperatures $k_BT_I\sim e^2/2a_0=1\text{Ry}$ and electron
densities such that $T_F\sim T_I$, i.e. $n\sim 0.048$ or $r_s\sim
1.7$. \cite{Pierleoni1994} Molecules may form at even lower
temperatures. Nonetheless in our stable purely fermionic mixture with
an equal species mass we see, already at the chosen thermodynamic
state, the unlike species binding and a spontaneous symmetry breaking
for like species bindings in a demixing transition
\cite{Fantoni11e,Fantoni13d,Fantoni14a}.   

We intend to adopt this method to simulate the two component plasma in
a curved surface \cite{Fantoni03jsp,Fantoni2008,Fantoni2012,Fantoni2012b}
in the near future. For example it could be interesting to study the
two component plasma on the surface of a sphere with a magnetic
monopole at the center \cite{Melik2001}.

%\appendix
%%%%%%%%%%%%%%%%%%%%%%%%%%%%%%%%%%%%%%%%%%%%%%%%%%%%%%%%%%%%%%%%%%%%%%%%%%%%%%%
%\section{...}
%%%%%%%%%%%%%%%%%%%%%%%%%%%%%%%%%%%%%%%%%%%%%%%%%%%%%%%%%%%%%%%%%%%%%%%%%%%%%%%
%\label{app:...}

%%%%%%%%%%%%%%%%%%%%%%%%%%%%%%%%%%%%%%%%%%%%%%%%%%%%%%%%%%%%%%%%%%%%%%%%%%%%%% 
%\begin{acknowledgments}

%\end{acknowledgments} 
%%%%%%%%%%%%%%%%%%%%%%%%%%%%%%%%%%%%%%%%%%%%%%%%%%%%%%%%%%%%%%%%%%%%%%%%%%%%%%
%\bibliographystyle{}
\bibliography{jft}

%%%%%%%%%%%%%%%%%%%%%%%%%%%%%%%%%%%%%%%%%%%%%%%%%%%%%%%%%%%%%%%%%%%%%%%%%%%%%%
%%%%%%%%%%%%%%%%%%%%%%%%%%%%%%%%%%%%%%%%%%%%%%%%%%%%%%%%%%%%%%%%%%%%%%%%%%%%%%
%%%%%%%%%%%%%%%%%%%%%%%%%%%%%%%%%%%%%%%%%%%%%%%%%%%%%%%%%%%%%%%%%%%%%%%%%%%%%%
\end{document}